\documentclass[aps,prb,twocolumn]{revtex4-1}
\usepackage{amssymb}
\usepackage{graphicx}

\begin{document}
\author{Amit, R. K. Gopal, and Yogesh Singh }
\affiliation{Department of Physical Sciences, Indian Institute of Science Education and Research (IISER) Mohali, Knowledge City, Sector 81, Mohali 140306, India.}

\date{\today}

\title{Observation of Chiral character deep in the topological insulating regime in Bi$_{1-x}$Sb$_x$}

\begin{abstract}
Bi$_{1-x}$Sb$_x$ is a topological insulator (TI) for $x \approx 0.03 $--$0.20$.  Close to the Topological phase transition at $x = 0.03$, a magnetic field induced Weyl semi-metal (WSM) state is stabilized due to the splitting of the Dirac cone into two Weyl cones of opposite chirality.  A signature of the Weyl state is the observation of a Chiral anomaly [negative longitudnal magnetoresistance (LMR)] and a violation of the Ohm's law (non-linear $I-V$).  We report the unexpected discovery of a Chiral anomaly in the whole range ($x = 0.032, 0.072, 0.16$) of the TI state.  This points to a field induced WSM state in an extended $x$ range and not just near the topological transition at $x = 0.03$. Surprisingly, the strongest Weyl phase is found at $x = 0.16$ with a non-saturating negative LMR much larger than observed for $x = 0.03$.  The negative LMR vanishes rapidly with increasing angle between $B$ and $I$.  Additionally, non-linear $I$--$V$ is found for $x = 0.16$ indicating a violation of Ohm's law.  This unexpected observation of a strong Weyl state in the whole TI regime in Bi$_{1-x}$Sb$_x$ points to a gap in our understanding of the detailed electronic structure evolution in this alloy system.   
\end{abstract}

\maketitle  

Condensed matter physics has taken an exciting turn after the discovery of two and three dimensional (2D/3D) topological materials possessing a nontrivial bulk band structure characterized by topological invariants [\onlinecite{Hasan2010,Kane2005,Moore2010,Ando2013}]. In particular, Topological (Dirac or Weyl) semi-metals (DSM, WSM), which possess bulk bands with linear electronic dispersions in all three momentum directions, have attracted immense recent interest [\onlinecite{Wan2011,Young2012, Burkov2016, Borisenko2014, Liu2014, Lv2015, Lu2015, Huang2015, Weng2015, Liu2014b}].  In DSMs, doubly degenerate Dirac cones (of opposite chirality) exist which are protected by time reversal (TRS) and inversion (IS) symmetry.  Na$_3$Bi [\onlinecite{Liu2014}] and Cd$_3$As$_2$ [\onlinecite{Liu2014b}] are examples of the most studied DSM materials.  Breaking of any of the above symmetries leads to a splitting of the Dirac cones into a pair of Weyl cones of opposite chirality, turning the DSM into a WSM.  A WSM state has been shown for example, in the mono-arsenides TaAs and NbAs where the IS is broken.  A TRS broken WSM state has been found for example in YbMnBi$_2$ [\onlinecite{Borisenko2015}].  Another route to designing a WSM has been demonstrated for the half-Heuslar material GdPtBi which, in zero magnetic fields, is a gapless semiconductor with quadratic bands. On application of a magnetic field, the quadratic bands are Zeeman split and Weyl nodes are formed at the crossings of the Zeeman split bands [\onlinecite{Hirschberger2016}].       

WSMs show several anomalous and potentially technologically useful transport properties such as very large charge carrier mobility $\mu$ (and hence a large electronic mean free path $l_e$), low carrier density, giant linear magnetoresistance (MR), anomalous Hall effect, and strong anisotropy in the MR with a negative longitudnal MR (LMR) when the magnetic field $B$ is applied parallel to the current $I$ [\onlinecite{Burkov2016, Hosur2013, Zhang2011, Wang2013, Abrikosov1998, Burkov2014, Burkov2015}].  
The negative LMR is a consequence of the Adler-Bell-Jackiw Chiral anomaly [\onlinecite{Adler1969, Bell1969}] predicted for Weyl Fermions in parallel magnetic and electric fields $B || E$ [\onlinecite{Nielsen1983}].  A crucial signature of the Chiral anomaly is that the negative LMR is extremely sensitive to the angle between $B$ and $E$ and is rapidly suppressed as the angle between $B$ and $E$ is increased.  The Chiral anomaly has been observed in the WSMs TaAs [\onlinecite{Zhang2015}], Na$_3$Bi [\onlinecite{Xiong2015}], Cd$_3$As$_2$ [\onlinecite{Li2015, Li2016}], and GdPtBi [\onlinecite{Hirschberger2016}] among others.     

The alloy system Bi$_{1-x}$Sb$_x$ has also been a very fruitful playground for the observation of various Topological states.  In fact the first 3D Topological Insulator state was experimentally verified in Bi$_{1-x}$Sb$_x$.  Bi is a trivial band insulator.  As Sb is introduced, a band inversion occurs at $x = 0.03$ signalling a Topological phase transition into a TI state beyond $x = 0.03$. The TI phase spans a large range $x = 0.03$--$0.22$ in the Bi$_{1-x}$Sb$_x$ system [\onlinecite{Fu2007, Teo2008, Guo2011, Hsieh2009, Teo2009, Hsieh2008}].  %In the composition range $x = 0.03$ -- $0.22$, ARPES and STM/STS studies have revealed multiple odd number of band crossings close to the narrow bulk band gap [\onlinecite{Hsieh2009, Teo2009, Hsieh2008}].  

At the Topological transition point $x = 0.03$, the material is a DSM which changes to a WSM state on the application of a magnetic field.  Indeed the Chiral anomaly in the $B || E$ configuration with a strongly angle dependent negative LMR  has been observed for $x = 0.03$ material [\onlinecite{Kim2013}].  Additionally, a violation of the Ohm's law has been observed for $x = 0.05$ material which also lie close to the Topological phase transition into a TI state, and has been argued to be a consequence of the Chiral anomaly [\onlinecite{Shin2017}].  This study also concluded that no negative LMR was observed for samples far away $x > 0.05$ from the Topological transition [\onlinecite{Shin2017}].  Thus, Bi$_{1-x}$Sb$_x$ alloys host a TI state in an extended range $x = 0.03$ -- $0.22$, a DSM state at $x = 0.03$, and a WSM state for $x = 0.03$ when TRS is broken by the application of a magnetic field.  Very recently, the Sb rich side of Bi$_{1-x}$Sb$_x$ has been investigated theoretically and two new WSM phases have been predicted at large Sb doping of $x = 0.5$ and $x = 0.83$  [\onlinecite{Su2018}].  This suggests that new surprises are yet to be discovered in this well studied system.

In this article we have made a detailed magneto-transport study on six high quality single crystals of Bi$_{1-x}$Sb$_x$, covering the whole TI range $x = 0.032$--$0.16$.  We provide clear evidence that in addition to the crystal at $x = 0.032$ which is at the Topological phase transition, samples at $x = 0.072$ and $0.16$ also show the Chiral anomaly with a negative LMR which is strongly suppressed on increasing the angle between $B$ and $I$.  Additionally, non-linear $I$--$V$ is observed for $x = 0.16$ crystals only when $B || I$.  The negative LMR for the $x = 0.072$ sample is weaker than that for $x = 0.032$ and $x = 0.16$ samples.  The negative LMR for $x = 0.16$ is the strongest.  These observations strongly indicate that a WSM state exists for other compositions $x$ even far away from the Topological phase transition at $x = 0.03$.  These are unexpected results given that beyond $x \approx 0.04$, the Bi$_{1-x}$Sb$_x$ system is considered to be a TI with no bulk bands.  Our discovery calls for a careful relook at existing ARPES and STM data to identify the exact locations of the discovered Weyl nodes.

\noindent
\emph{Experimental:}
Single crystals of Bi$_{1-x}$Sb$_x$  ($0\leq x \leq.16$) were grown using a modified Bridgeman technique.  Stoichiometric amounts of Bi (5N) and Sb (5N) shots were sealed in a quartz tube under vacuum.  The quartz tube was placed vertically in a box furnace and heated to $650~^o$C in $15$~hrs, kept there for $8$~hrs, and then slowly cooled to $270~^o$C over a period of five days for crystal growth.  The crystals are kept at $270~^o$C for seven days for annealing and homogenization.  Large shiny crystals could be cleaved from the resulting boul.  The structure of the crystals was confirmed using powder x-ray diffraction (PXRD) on crushed crystals which was collected at room temperature using a Rikagu diffractometer (Cu K$\alpha$).   Chemical analysis of the crystals was done using energy dispersive spectroscopy (EDS) on a JEOL scanning electron microscope (SEM).  Electrical and magnetic transport was measured using a Quantum Design Physical Property Measurement System (PPMS).

\noindent
\emph{Results:}
 
\begin{figure}[t]   
\includegraphics[width= 3 in]{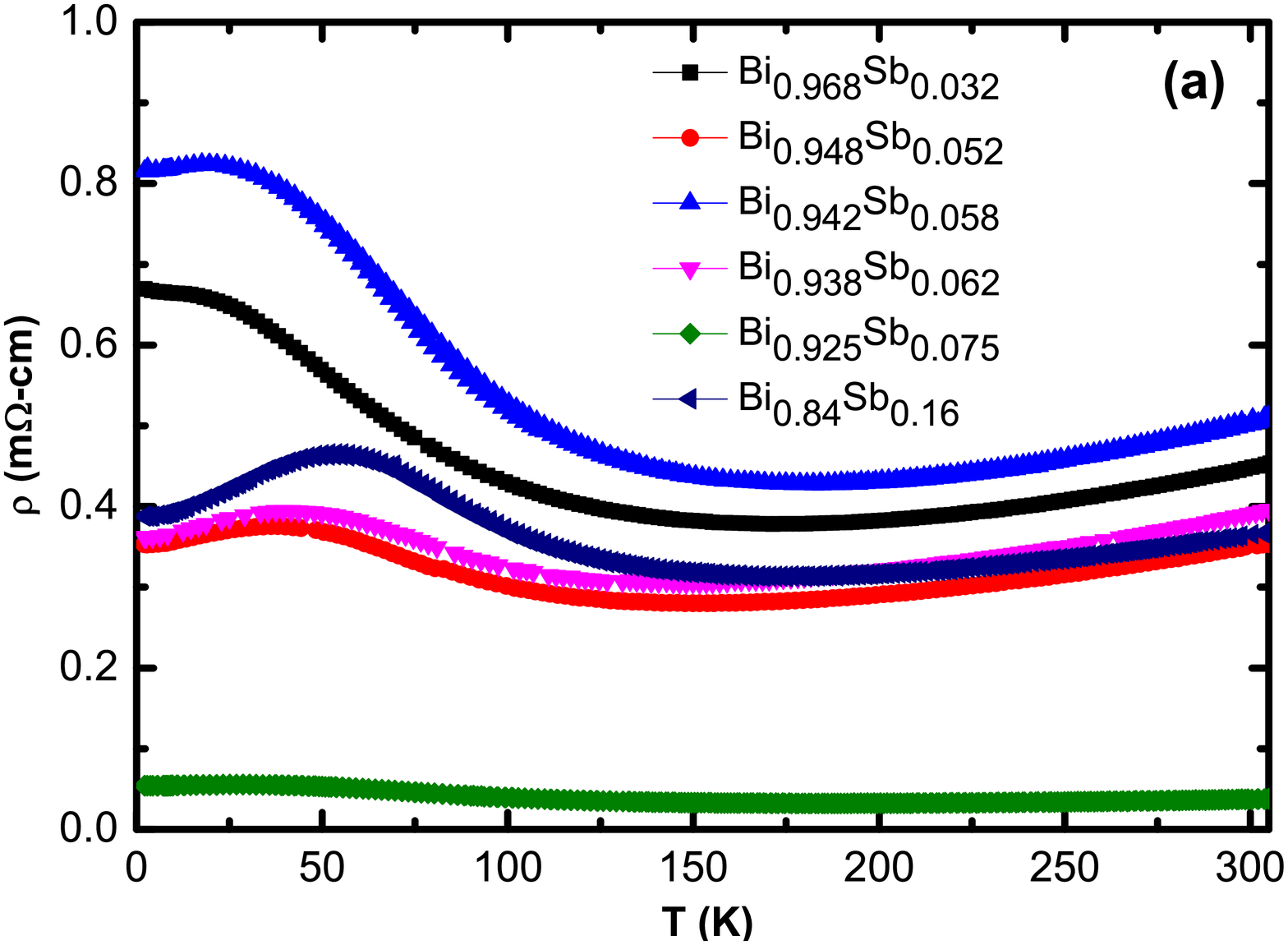}
\includegraphics[width= 3 in]{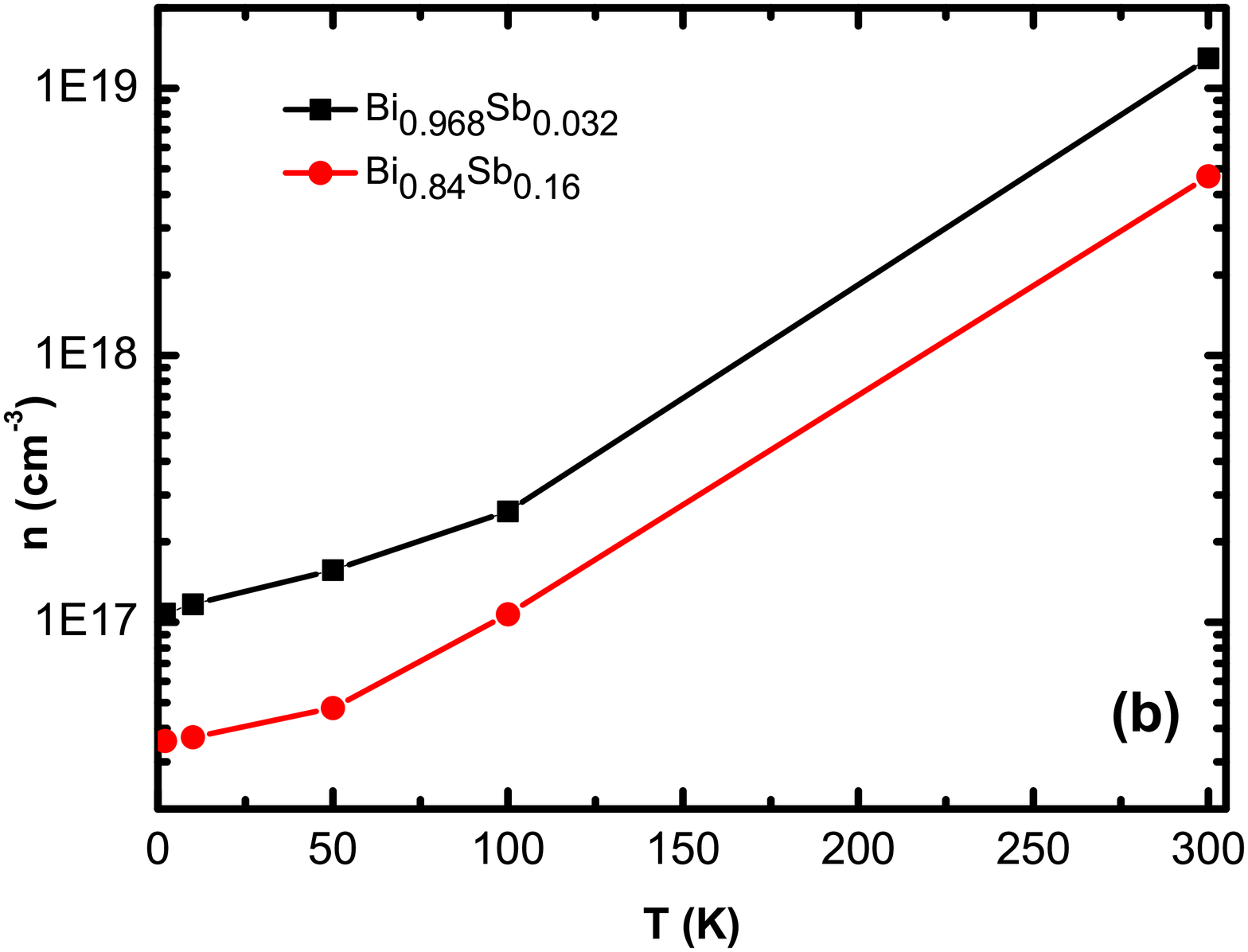}
\includegraphics[width= 3 in]{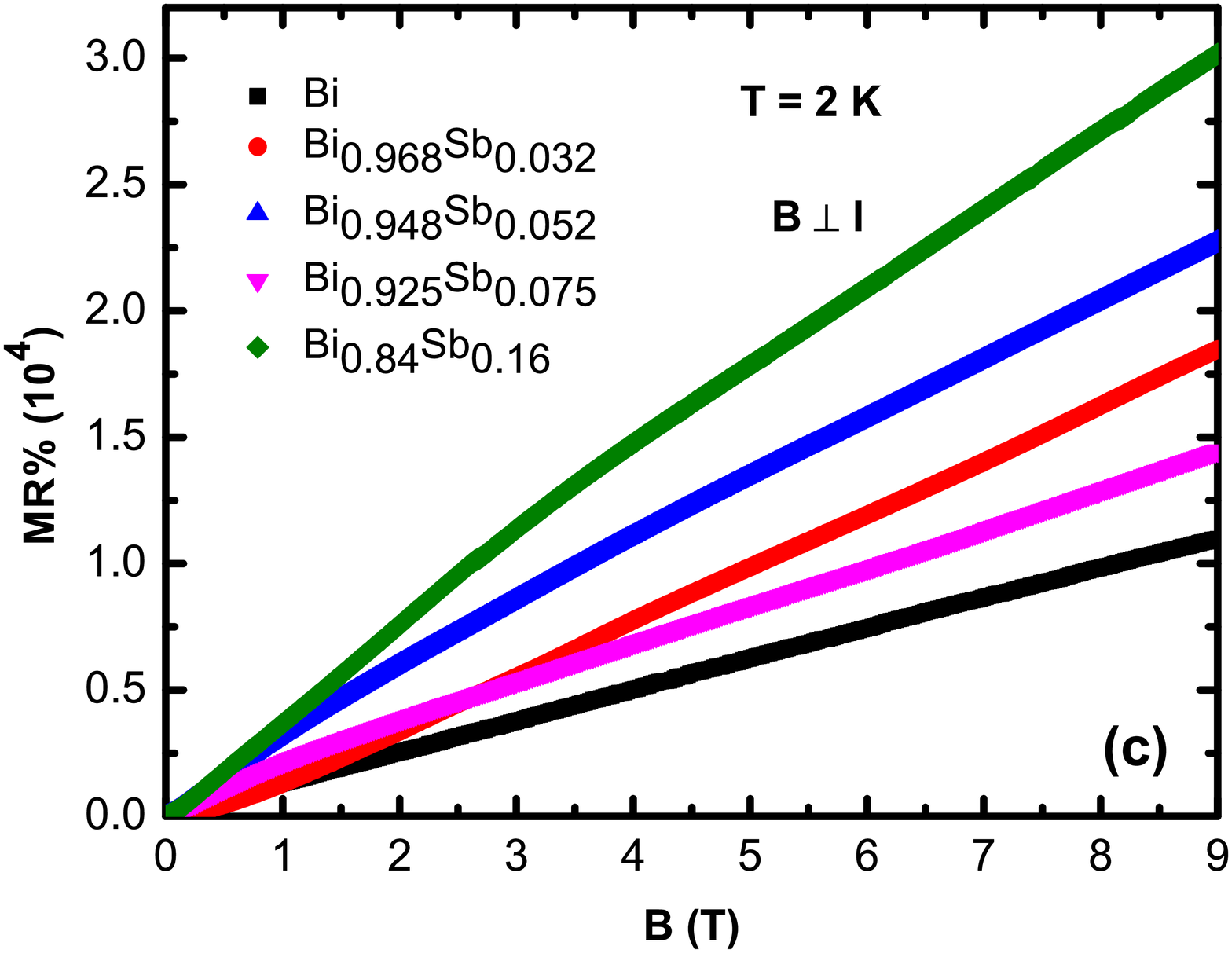}    
\caption{(Color online) (a) Resistivity $\rho$ vs temperature $T $ for Bi$_{1-x}$Sb$_x$.  (b) The carrier density $n$ versus $T$ for $x = 0.032, 0.16$ samples.  (c) Transverse magneto resistance $MR$ vs $B$ at $T = 2$~K for Bi$_{1-x}$Sb$_x$.  The $MR$ for all $x$ is positive and linear at high $B$.   
\label{Fig-R}}
\end{figure} 

We first show that our Bi$_{1-x}$Sb$_x$  ($0.032\leq x \leq.16$) crystals show the expected conventional transport properties reported previously.  Figure~\ref{Fig-R}~(a) shows the $\rho(T)$ data for Bi$_{1-x}$Sb$_x$ with $x \geq 0.032$.  For all these materials the qualitative $T$ dependence is similar.  There is a weak metallic behaviour at high temperatures before a minimum in $\rho(T)$ is observed around $T = 100$--$150$~K\@.  For lower $T$ the $\rho(T)$ increases as $T$ decreases.  Finally the $\rho(T)$ passes over a maximum and shows metallic behaviour for lower temperatures.  Such a $\rho(T)$ is expected for a topological insulator where the charge gap in the bulk band structure suppresses bulk conduction below some $T$ and $\rho$ increases below this $T$.  For sufficiently low $T$ where the bulk conductivity becomes too small, conduction through surface states takes over and $\rho(T)$ stops increasing.  The competition between surface and bulk conductivity influences the position of maxima and minima in $\rho(T)$.  Figure~\ref{Fig-R}~(b) shows the carrier density $n$ estimated from Hall measurements, as a function of $T$ for the $x = 0.032, 0.16$ samples.  The $n$ drops by about two orders of magnitude on cooling from $300$~K to $2$~K indicating the gapped nature of the crystals.

Large conventional transverse magnetoresistance with magnetic field perpendicular to the current direction has been reported in the TI state.  Transverse magnetoresistance (MR) for Bi$_{1-x}$Sb$_x$  with $B \perp I$ is shown in Fig.~\ref{Fig-R}~(c).  We observe large non-saturating positive MR for all samples which is linear up to the largest $B$, consistent with previous observations on samples close to $x = 0.04$.   Thus our samples show behaviour consistent with them being in the TI state.  We note that the MR is largest for the $x = 0.16$ sample and is much larger than reported previously for the $x = 0.04$ samples close to the  trivial to Topological insulator transition. 

\begin{figure}[t]   
\includegraphics[width= 3 in]{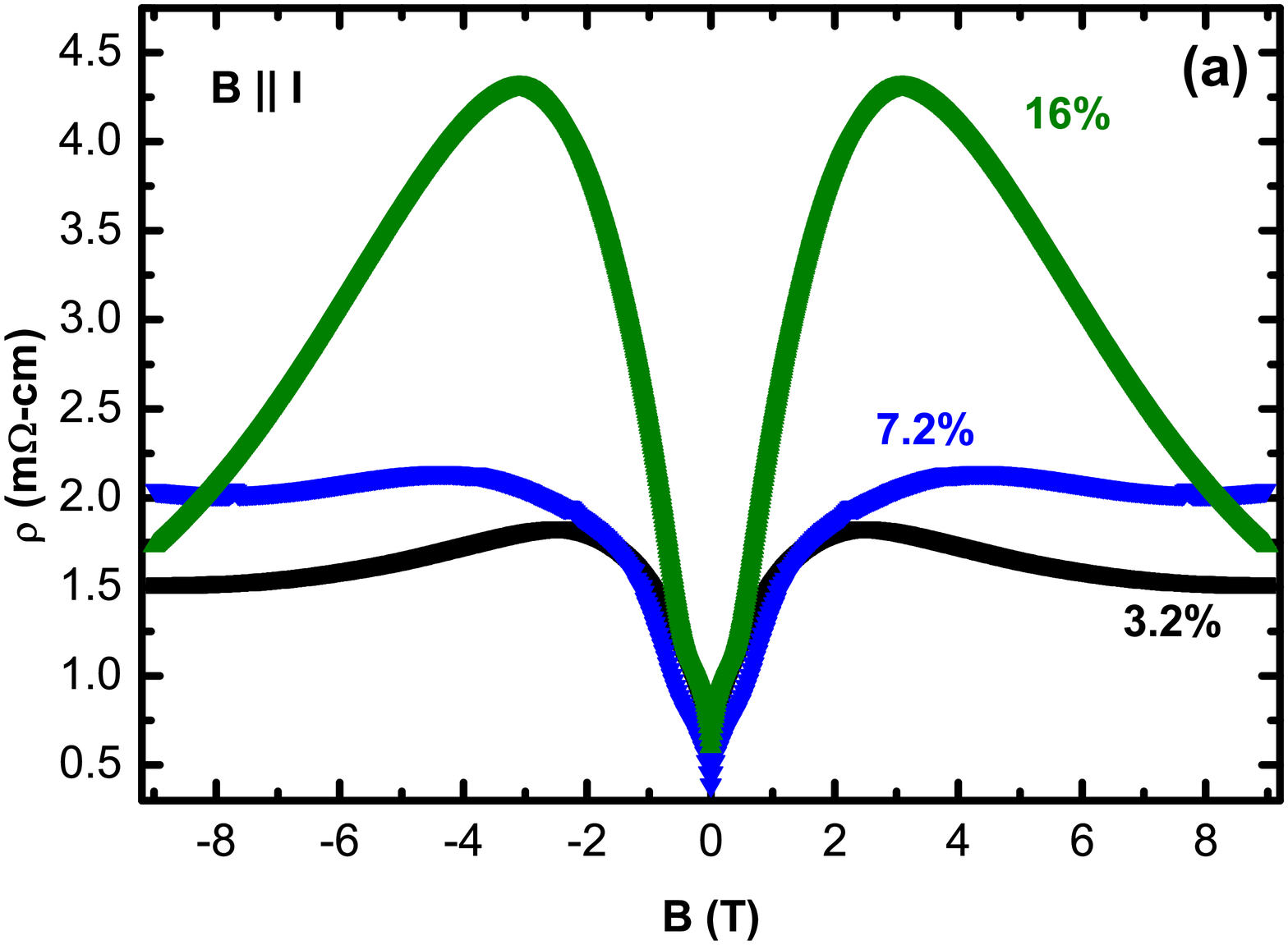}
\includegraphics[width= 3 in]{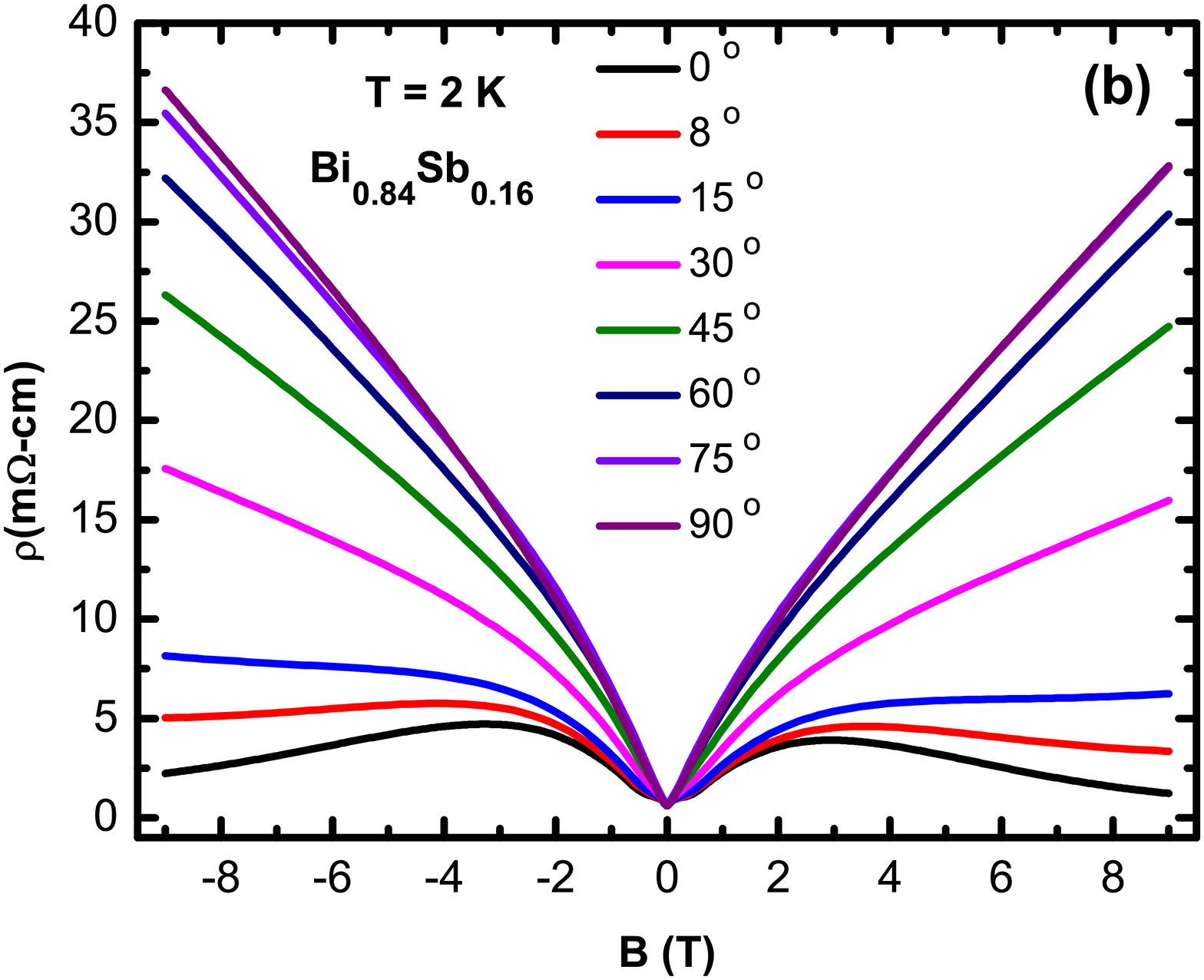}        
\caption{(Color online) (a) Resistivity $\rho$ vs magnetic field $B$ at $T = 2$~K with $B || I$ for Bi$_{1-x}$Sb$_x$ ($x = 3.2, 7.2, 16$~\%).  (b) $\rho$ vs $B$ at $T = 2$~K for the $x = 16$\%  sample with $B$ applied at various angles to the current $I$.  The negative LMR seen clearly for $B || I$ is rapidly suppressed as the angle between $B$ and $I$ increases.     
\label{Fig-LMR}}
\end{figure}

We now show evidence for the Chiral anomaly in the whole TI region.  Figure~\ref{Fig-LMR}~(a) shows the resistivity $\rho$ vs magnetic field $B$ measured at temperature $T = 2$~K with the magnetic field $B$ applied parallel to the electrical current $I$ for Bi$_{1-x}$Sb$_x$ ($x = 3.2, 7.2, 16$\%).   These three samples are located at (i) the transition from the normal insulator to a TI ($x = 0.032$), (ii) in the middle of the $x$ range in which a TI has been shown to exist ($x = 0.072$), and (iii) near the end of the range of the TI state ($x = 0.16$).  For the $x = 0.032$ sample, we observe that after the initial increase in $\rho$ at small $B$ most likely arising from weak anti-localization (WAL), the $\rho$ turns down and starts decreasing upto the highest $B$ measured.  This is the negative longitudnal ($B || I$) magnetoresistance (NLMR) or the Chiral anomaly.  For $x = 0.032$ the Bi$_{1-x}$Sb$_x$ is situated close to the Dirac semi-metal state and the application of a magnetic field leads to time-reversal symmetry breaking and hence the Dirac cone is expected to split into a pair of Weyl nodes and so the Chiral anomaly (NLMR) for this composition is expected and has been observed previously as well for Bi$_{0.096}$Sb$_{0.04}$ \cite{Kim2013} and Bi$_{0.095}$Sb$_{0.05}$ \cite{Shin2017} samples.  

What is surprising is that we observe a Chiral anomaly even for samples far away from $x = 0.032$.  Figure~\ref{Fig-LMR}~(a) also shows the $\rho$ vs $B$ data for $x = 0.072$ and $x = 0.16$.  The $x = 0.072$ sample shows a negative LMR above $B \sim 4$~T suggesting a Weyl state for this $x$ too.  However, the $\rho$ data for $x = 0.16$ shows the strongest negative LMR compared to even the $x = 0.032$ sample.  The NLMR keeps increasing up to the highest magnetic fields measured $B = 9$~T\@.  This strongly suggests that a WSM state exists for $x = 0.16$ as well.   %We were able to fit our LMR data shown in Fig.~\ref{Fig-LMR}~(a) using the expression $n$

A crucial signature of the Chiral anomaly is the strong dependence of the NLMR on the angle between $B$ and $I$.  Figure~\ref{Fig-LMR}~(b) shows the angle dependence of the NLMR for the $x = 0.16$ sample.  We observe that for angle $ = 0$ ($B || I$), the NLMR is the largest and as the angle between $B$ and $I$ is increased, the magnetoresistance quickly increases and changes to completely positive MR for angle $\geq 8~^o$.  This strong sensitivity of the NLMR to the angle between $B$ and $I$ is strong evidence of the Chiral anomaly for the $x = 0.16$ sample and points to it being in the WSM state.         

\begin{figure}[t]   
\includegraphics[width= 3 in]{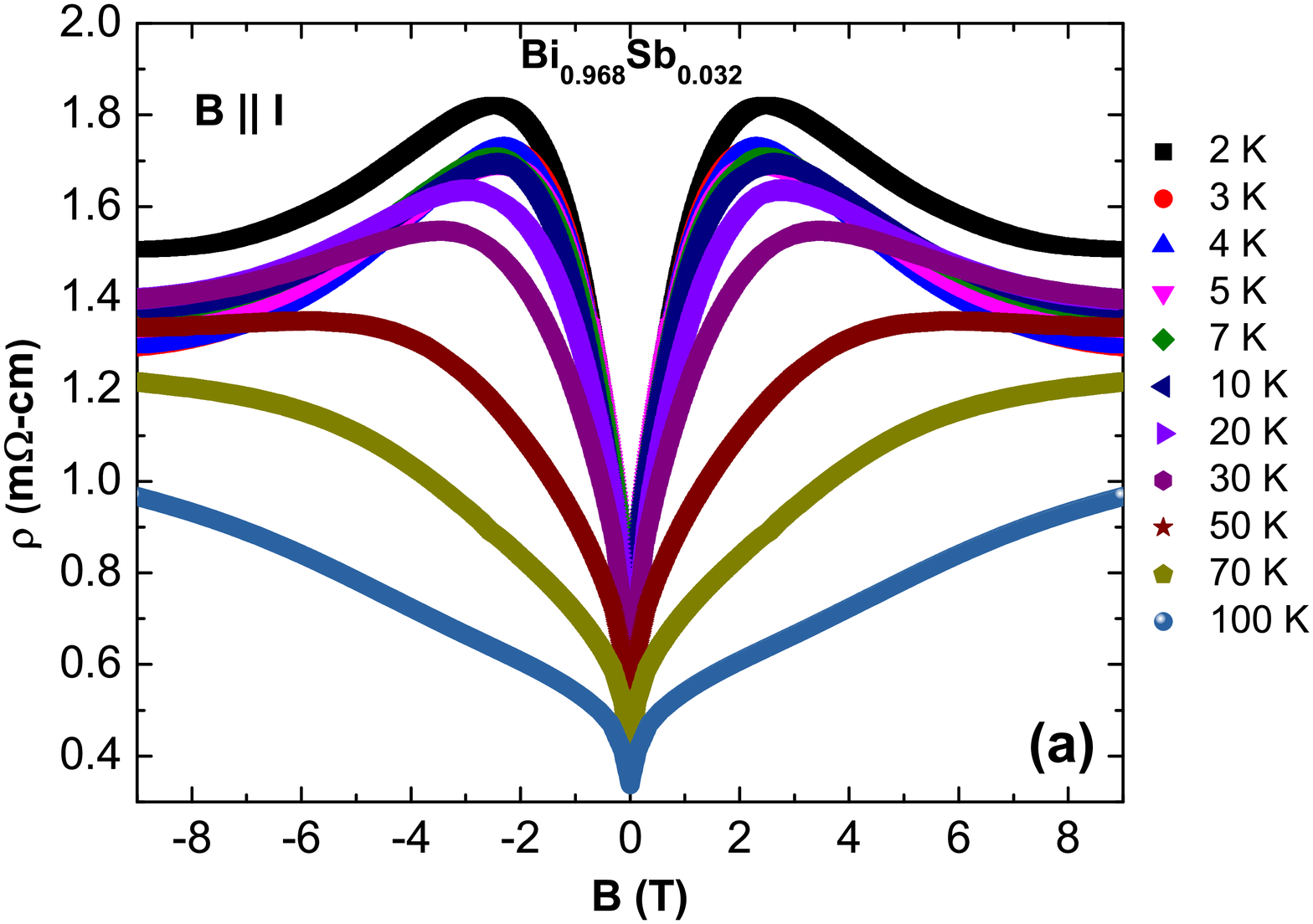}
\includegraphics[width= 3 in]{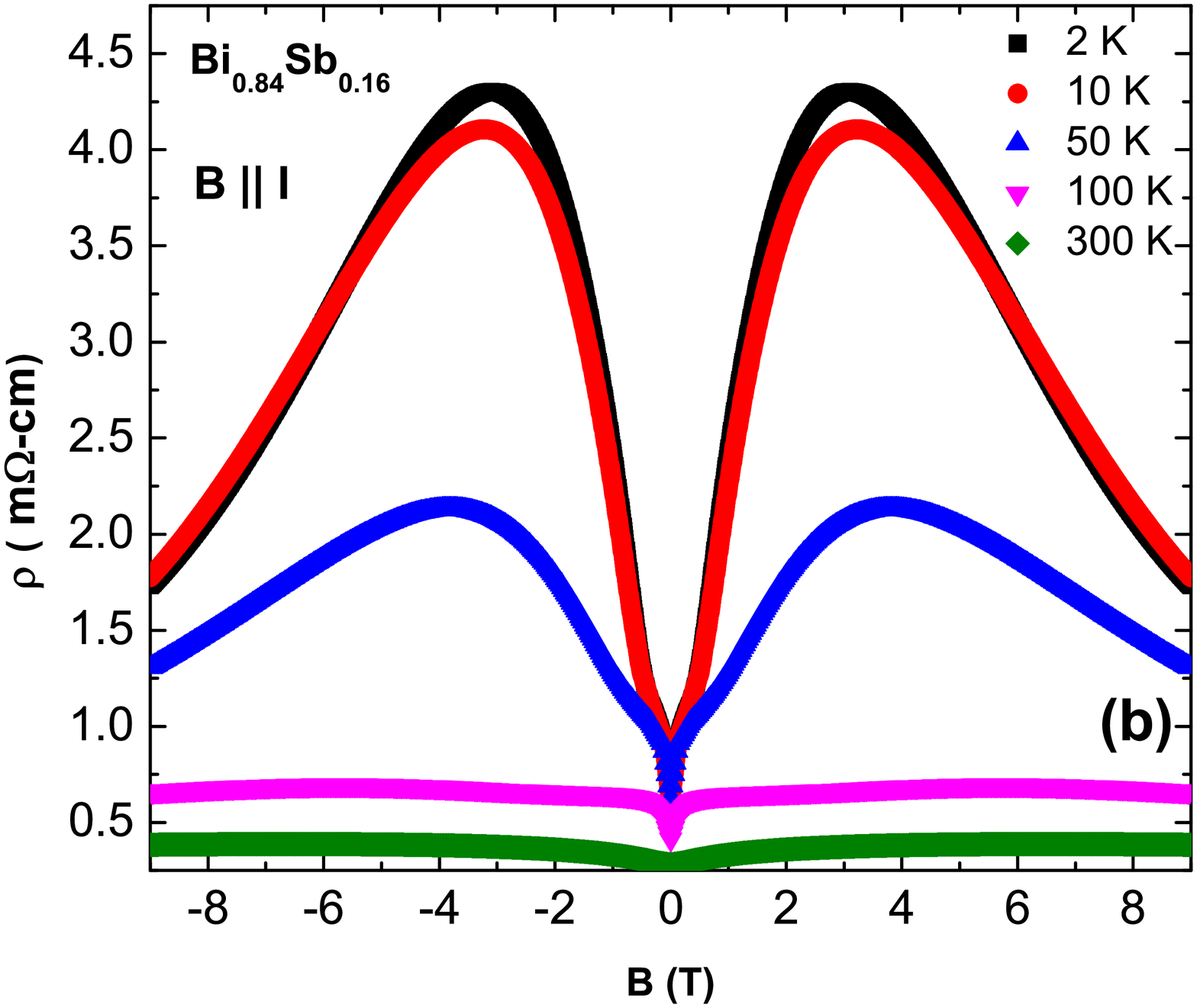}        
\caption{(Color online) Temperature dependence of the negative LMR for the (a) $x = 0.032$ and (b) $x = 0.16$ samples.         
\label{Fig-LMR-T}}
\end{figure}

We have also measured the $T$ dependence of the LMR for the $x = 3.2$ and $16\%$ samples as shown in Fig.~\ref{Fig-LMR-T}~(a) and (b), respectively.  The negative contribution to the LMR shows a strong $T$ dependence and is suppressed for higher $T$.  The NLMR is no longer observed for $T > 50$~K for $x = 0.032$ and $x = 0.16$.  

\begin{figure}[t]   
\includegraphics[width= 3.4 in]{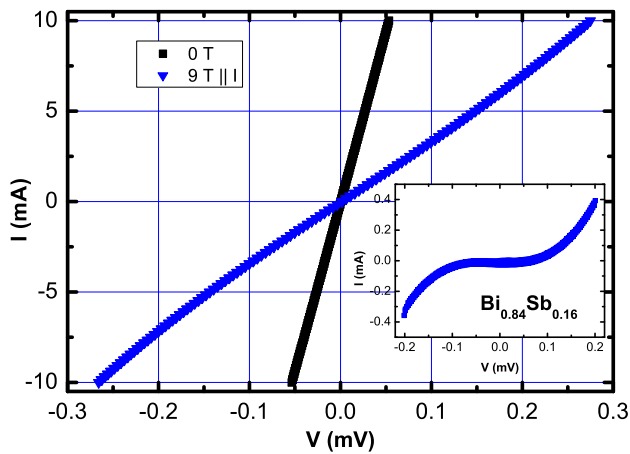}
\caption{(Color online) $I$--$V$ curve for the $x = 0.16$ sample measured at $T = 2$~K in $B = 0$ and in $B = 9$~T for $B || I$.      
\label{Fig-IV}}
\end{figure}

Further confirmation of anomalous transport behaviour expected for a WSM is obtained for the $x = 0.16$ sample from the observation of a violation of the Ohm's law.  Figure~\ref{Fig-IV} shows the $I$--$V$ curve for the $x = 0.16$ sample measured at $T = 2$~K in $B = 0$ and in $B = 9$~T for $B || I$.  The $B = 0$ data are completely linear as expected in conventional metals.  The $I$--$V$ data at $B = 9$~T with $B \perp I$ were also measured (not shown) and were found to be completely linear.  For $B || I$, the LMR configuration, a clear non-linearity can be seen in the $I$--$V$ curves.  To highlight the non-linearity, we have subtracted the linear part of the $I$--$V$ curve obtained by a fit to the $I \leq 2$~mA data.  The resulting $I$--$V$ curve obtained at $9$~T is shown in Fig.~\ref{Fig-IV}~inset and clearly shows a non-linear behaviour.  This clear violation of the Ohm's law has previously been reported only for the $x = 0.05$ samples close to the Topological transition boundary \cite{Shin2017}. \\

\noindent
\emph{Summary and Discussion:} We have studied the magneto-transport properties of Bi$_{1-x}$Sb$_x$ single crystals for $x = 0.032$--$0.16$ spanning almost the whole range of Sb substitution for which Bi$_{1-x}$Sb$_x$ has previously been shown to be a Topological Insulator (TI).  The transition from trivial insulator to a TI occurs at $x \approx 0.03$.  Samples at this topological phase transition are Dirac semi-metals (DSM) and the application of a magnetic field breaks time reversal symmetry and hence splits the Dirac cone into two opposite chirality Weyl cones displaced in momentum space along the field direction turning the material into a Weyl semi-metal (WSM).  A smoking gun signature of a WSM state is the observation of a Chiral anomaly i.e. a negative longitudnal magnetoresistance (LMR) when a magnetic field $B$ is applied along the current $I$ direction.  This negative LMR is expected to be strongly suppressed as the angle between $B$ and $I$ is increased.  Additionally, a non-linear $I$--$V$ has been reported for the WSM state at $x = 0.05$.  We indeed observe the Chiral anomaly for our $x = 0.03$ crystals in agreement with previous reports on samples with $x = 0.04$ and $0.05$.  Unexpectedly, we also find strong evidence of a WSM state for the $x = 0.16$ sample, which is close to the end of the TI state.  In particular, we find a negative LMR even stronger than the $x = 0.032$ sample. The negative LMR is strongly suppressed with increasing angle between $B$ and $I$.  Additionally we observe a non-linear $I$-$V$ for $B || I$ but not for $B \perp I$.  A weaker but clearly observable negative LMR is also found for the $x = 0.072$ sample.  These observations strongly indicate that in addition to $x = 0.03$, a WSM state exists for $x = 0.072, 0.16$ materials expected to be deep in the TI state.    

Previous STM and ARPES results have shown a complex band structure in the bulk of these alloys for the compositions studied here.  Our results suggest that these bulk bands host Weyl nodes which are responsible for the observation of the anomalous longitudnal magneto-transport properties including the Chiral anomaly and violation of Ohm's law.  \\  

\noindent
\emph{Acknowledgments.--} We thank the X-ray facility at IISER Mohali.


\begin{references}

\bibitem{Hasan2010} M. Z. Hassan and C. Kane, Rev. Mod. Phys. {\bf 82}, 3045 (2010).
	
\bibitem{Kane2005} C. L. Kane and E. J. Mele, Phys. Rev. Lett. \textbf{95}, 146802 (2005).

\bibitem{Moore2010} J. E. Moore, Nature {\bf 464}, 194 (2010).

\bibitem{Ando2013} Y. Ando, J. Phys. Soc. Japan {\bf 82}, 102001 (2013).

\bibitem{Wan2011} X. Wan, A. M. Turner, A. Vishwanath, and S. Y. Savrasov, Phys. Rev. B {\bf 83}, 205101 (2011).

\bibitem{Young2012} S. M. Young, S. Zaheer, J. C. Y. Teo, C. L. Kane, E. J. Mele, and A. M. Rappe, Phys. Rev. Lett. {\bf 108},140405 (2012). 

\bibitem{Burkov2016} A. A. Burkov, Nature Materials {\bf 15}, 1145 (2016).

\bibitem{Borisenko2014} S. Borisenko, Q. Gibson, D.l Evtushinsky, V. Zabolotnyy, B. Büchner, and R. J. Cava, Phys. Rev. Lett. {\bf 113}, 027603 (2014).

\bibitem{Liu2014} Z. K. Liu, B. Zhou, Y. Zhang, Z. J. Wang, H. M. Weng, D. Prabhakaran, S.-K. Mo, Z. X. Shen, Z. Fang, X. Dai, Z. Hussain, Y. L. Chen, Science {\bf 343}, 864 (2014).

\bibitem{Lv2015} B.Q. Lv, H.M. Weng, B.B. Fu, X.P. Wang, H. Miao, J. Ma, P. Richard, X.C. Huang, L.X. Zhao, G.F. Chen, Z. Fang, X. Dai, T. Qian, and H. Ding, Phys. Rev. X {\bf 5}, 031013 (2015).

\bibitem{Lu2015} L. Lu, Z. Wang, D. Ye, L. Ran, L. Fu, J. D. Joannopoulos, M. Soljacic Science {\bf 349}, 622 (2015).

 \bibitem{Huang2015}  S-M. Huang, S-Y. Xu, I. Belopolski, C-C. Lee, G. Chang, B. Wang, N. Alidoust, G. Bian, M. Neupane, C. Zhang, S. Jia, A. Bansil, H. Lin, and M. Z. Hasan, Nat. Commun. {\bf 6}, 7373 (2015)
 
 \bibitem{Weng2015} H. Weng, C. Fang, Z. Fang, B. A. Bernevig, and X. Dai, Phys. Rev. X {\bf 5}, 011029 (2015).
 
 \bibitem{Liu2014b} Z. K. Liu, J. Jiang, B. Zhou, Z. J. Wang, Y. Zhang, H. M. Weng, D. Prabhakaran, S-K. Mo, H. Peng, P. Dudin, T. Kim, M. Hoesch, Z. Fang, X. Dai, Z. X. Shen, D. L. Feng, Z. Hussain, and Y. L. Chen, Nat. Mat. {\bf 13}, 677 (2014).

\bibitem{Borisenko2015} S. Borisenko, D. Evtushinsky, Q. Gibson, A. Yaresko, T. Kim, M. N. Ali, B. Buechner, M. Hoesch, and R. J. Cava, arXiv:1507.04847 (2015).

\bibitem{Hirschberger2016} M. Hirschberger, S. Kushwaha, Z. Wang, Q. Gibson, S. Liang, C. A. Belvin, B. A. Bernevig, R. J. Cava, and N. P. Ong, Nat. Mat. {\bf 15}, 1161 (2016).

\bibitem{Hosur2013} P. Hosur and X. Qi, Comptes Rendus Physique {\bf 14}, 857 (2013).

\bibitem{Zhang2011} W. Zhang, Y. Rui, F. Wanxiang, Y. Yugui, W. Hongming, D. Xi, and F. Zhong, Phys. Rev. Lett. {\bf 106}, 156808 (2011).

\bibitem{Wang2013} Z. Wang, H. Weng, Q. Wu, X. Dai, and Z. Fang, Phys. Rev. B {\bf 88}, 125427 (2013).

\bibitem{Abrikosov1998} A. A. Abrikosov, Phys. Rev. B {\bf 58}, 2788 (1998).

\bibitem{Burkov2014} A. A. Burkov, Phys. Rev. Lett. {\bf 113}, 187202 (2014).

\bibitem{Burkov2015} A. A. Burkov, J. Phys.:Condens. Matter. {\bf 27}, 113201 (2015).

\bibitem{Adler1969} S. L. Adler, Phys. Rev. 177, 2426?2438 (1969).

\bibitem{Bell1969} J. S. Bell, R. Jackiw, Nuovo Cim. 60A, 4 (1969).

\bibitem{Nielsen1983} H. B. Nielsen, M. Ninomiya, Phys. Lett. B 130, 389?396 (1983).

\bibitem{Zhang2015} C. Zhang, S.-Y. Xu, I. Belopolski, Z. Yuan, Z. Lin, B. Tong, N. Alidoust, C.-C. Lee, S.-M. Huang, H. Lin, M. Neupane, D. S. Sanchez, H. Zheng, G. Bian, J. Wang, C. Zhang, T. Neupert, M. Z. Hasan, S. Jia, Nat. Commun. {\bf 7}, 10735 (2015).

\bibitem{Xiong2015} J. Xiong, S. K. Kushwaha, T. Liang, J. W. Krizan, M.Hirschberger, W. Wang, R. J. Cava, and N. P. Ong, ?Evidence for the chiral anomaly in the Dirac semimetal Na3 Bi,? Science 350,413 (2015).

\bibitem{Li2015} C-Z. Li, L-X. Wang, H. Liu, J. Wang, Z-M. Liao, and D-P. Yu, Nat. Commun. {\bf 6}, 10137 (2015).

\bibitem{Li2016} H. Li, H. He, H-Z. Lu, H. Zhang, H. Liu, R. Ma, Z. Fan, S-Q. Shen, and J. Wang, Nat. Commun. {\bf 7}, 10301 (2016).

\bibitem{Fu2007} L. Fu and C. L. Kane, Phys. Rev. B {\bf 76}, 045302 (2007). 

\bibitem{Teo2008} J. C. Y. Teo, L. Fu, and C. L. Kane, Phys. Rev. B {\bf 78}, 045426 (2008).

\bibitem{Guo2011} H. Guo, K. Sugawara, A. Takayama, S. Souma, T. Sato, N. Satoh, A. Ohnishi, M. Kitaura, M. Sasaki, Q.-K. Xue, and T. Takahashi, Phys. Rev. B {\bf 83}, 201104(R) (2011).

\bibitem{Hsieh2009} D. Hsieh, Y. Xia, L. Wray, D. Qian, A. Pal, J. H. Dil, J. Osterwalder, F. Meier, G. Bihlmayer, C. L. Kane, Y. S. Hor, R. J. Cava, and M. Z. Hasan, Science {\bf 323}, 919 (2009).

\bibitem{Teo2009} J. C. Y. Teo, L. Fu, and C. L. Kane, Phys. Rev. B 78, 45426 (2008).

\bibitem{Hsieh2008} D. Hsieh, D. Qian, L. Wray, Y. Xia, Y. S. Hor, R. J. Cava, and M. Z. Hasan. Nature {\bf 452}, 970 (2008).

\bibitem{Kim2013} H.-J. Kim, K.-S. Kim, J.-F. Wang, M. Sasaki, N. Satoh, A. Ohnishi, M. Kitaura, M. Yang, and L. Li, Phys. Rev. Lett. {\bf 111}, 246603 (2013).

\bibitem{Shin2017} D. Shin, Y. Lee, M. Sasaki, Y. H. Jeong, F. Weickert, J. B. Betts, H-J. Kim, K-S. Kim, and J. Kim, Nat. Mat. {\bf 16}, 1096 (2017).

\bibitem{Su2018} Y-H. Su, W. Shi, C. Felser, and Y. Sun, arXiv:1802.00288 (2018).

\end{references}
\end{document}